\documentclass[manuscript]{emulateapj}
\usepackage{multirow}
\usepackage{url}
\usepackage{color}
\begin{document}

\title{Origin of Ne Emission Line of Very Luminous Soft X-ray Transient MAXI J0158$-$744}
\author{Yukari Ohtani\altaffilmark{1,3}, Mikio Morii\altaffilmark{2} and Toshikazu Shigeyama\altaffilmark{3}}
\altaffiltext{1}{Department of Astronomy, Graduate School of Science, University of Tokyo, Bunkyo-ku, Tokyo 113-0033, Japan}
\altaffiltext{2}{MAXI team, Institute of Physical and Chemical Research (RIKEN),
2-1 Hirosawa, Wako, Saitama 351-0198, Japan.}
\altaffiltext{3}{Research Center for the Early Universe, Graduate School of Science, University of Tokyo, Bunkyo-ku, Tokyo 113-0033, Japan}

\begin{abstract}
We investigate the mechanism to reproduce notable spectral features at the ignition phase of nova explosion observed for a super-Eddington X-ray transient source MAXI J0158$-$744 in the Small Magellanic Cloud. These are a strong Ne IX emission line at 0.92 keV with a large equivalent width of $0.32^{+0.21}_{-0.11}$ keV and the absence of Ne X line at 1.02 keV. In this paper, we calculate the radiative transfer using a Monte Carlo code, taking into account the line blanketing effect due to transitions of N, O, Ne, Mg and Al ions in an accelerating wind emanating from a white dwarf with a structure based on a spherically symmetric stationary model. We found that the strong Ne IX line can be reproduced if the mass fraction of Ne is enhanced to $10^{-3}$ or more and that of O is reduced to $\sim5\times10^{-9}$ or less and that the absence of other lines including Ne X ions at higher energies can be also reproduced by the line blanketing effect. This enhancement of the Ne mass fraction indicates that the ejecta are enriched by Ne dredged up from the surface of the white dwarf composed of O, Ne, and Mg rather than C and O, as already pointed out in the previous work. We argue that the CNO cycle driving this nova explosion converted most of C and O into N and thus reduced the O mass fraction.
\end{abstract}


\section{Introduction}
MAXI (Monitor of All-sky X-ray Image; \citealt{2009PASJ...61..999M}) discovered a luminous soft X-ray transient,
MAXI J0158$-$744, in the direction of the east edge of the Small Magellanic Cloud (SMC) on November 11, 2011 \citep{2011ATel.3756....1K}. The spectra of the initial X-ray flash are reproduced by blackbody models with the temperature of $\sim300$ eV and the photospheric radius of $\sim2,000$ km \citep{2013ApJ...779..118M}. The Swift follow-up observations, which started from 0.44 days after the MAXI discovery, revealed that the X-ray spectra, which are fitted by blackbody spectra at temperature $\sim0.1$ keV, were similar to those of novae in the super-soft source (SSS) phase \citep{2012ApJ...761...99L}. The optical spectra taken by SAAO, ESO, and SMARTS telescopes indicated the characteristics of a Be star at the distance of the SMC \citep{2012ApJ...761...99L,2013ApJ...779..118M}.
Thus, MAXI J0158$-$744 was recognized as a nova accompanied by a Be star.
The X-ray flash detected by MAXI GSC (Gas Slit Camera; \citealt{2011PASJ...63S.623M,2011PASJ...63S.635S}) and SSC (Solid-state Slit Camera; \citealt{2010PASJ...62.1371T,2011PASJ...63..397T})
lasted for $\sim1300$ s at least. The maximum luminosity was $\sim100$ times brighter than the Eddington limit of a solar mass object.

MAXI J0158$-$744 behaved differently from most novae known up to the present. While standard novae emit optical photons for at least several days before emitting soft X-ray photons during the SSS phase, the outburst of MAXI J0158$-$744 was first detected in the X-ray band, not in optical. \citet{2013ApJ...779..118M} reported that the unabsorbed X-ray luminosity was about $3\times10^{39}$ erg s$^{-1}$ at $t=8$ s ($=t_{8}$) in the energy range of 0.7$-$7.0 keV, assuming the distance of 60 kpc (Hereafter, the origin of time is referred to the MAXI trigger time.). The X-ray luminosity reached the peak of $L_{\rm p}\sim2 \times 10^{40}$ erg s$^{-1}$ at $t=220$ s ($=t_{220}$),
and then decreased to $L_{1296}\sim7 \times 10^{39}$ erg s$^{-1}$ at $t=1296$ s ($=t_{1296}$). Since these luminosities exceed the Eddington luminosity for a solar mass object by a factor of more than 10, a strong wind is expected to blow from the surface of the white dwarf (WD). This source is also characterized by a lack of the early optical phase and the very fast decay of the initial X-ray flash. The scan periods of the MAXI detectors constrain the  duration  $\Delta t$ of the event to be less than $1.10\times10^{4}$ s. If the energy emitted in the X-ray band is supplied by hydrogen burning, we can estimate the mass of the exhausted hydrogen fuel as $M_{\rm H}=L_{\rm p}\Delta t/0.007c^{2}\sim2\times10^{-8}M_{\odot}$, which is much smaller than those in usual novae ($\gtrsim10^{-7}M_{\odot}$; \citealt{1982ApJ...257..767F}). Because of this tiny mass, the wind emanating from the WD becomes transparent soon after the thermonuclear runaway (TNR) outbreak. Especially, the most remarkable feature is the X-ray spectrum taken by MAXI/SSC at $t=t_{1296}$, which exhibits a strong He-like Ne (Ne IX) emission line at the energy of 0.92 keV and no prominent line of H-like Ne (Ne X) at 1.02 keV \citep{2013ApJ...779..118M}. We adopt the results of the analysis by \citet{2013ApJ...779..118M} using blackbody fits, defining $R_{1296}=2,290$ km and $k_{\rm B}T_{1296}=0.33$ keV  (where $k_{\rm B}$ is the Boltzmann constant) as fiducial values of the photospheric radius and temperature at $t=t_{1296}$.

Though \citet{2012ApJ...761...99L} tried to explain the super-Eddington luminosity by a shock induced model, \citet{2013ApJ...779..118M} pointed out the difficulty in such a shock-induced model. The latter authors instead argued that the observed phenomenon corresponds to the fireball phase at the ignition of a nova explosion and that the luminosity is possibly explained by the convection on the surface of the WD, releasing a large amount of energy produced by the TNR during the first 100 s. They inferred that the He-like Ne line can be explained by the optically thin region around the photosphere. In their work,  optically thin emission component with a temperature below 0.3 keV (Mekal component in \citealt{2013ApJ...779..118M}) and exceptionally large Ne abundance of 10 solar or more was necessary to explain the strong He-like Ne line and the lack of the H-like Ne line. In this case, a large emission measure ($\sim10^{63}$ cm$^{-3}$) is necessary and might be incompatible with the assumption of optically thin emission.

In this paper, we investigate the possibility that photons scattered by N, O, Ne, Mg, and Al ions in a supersonic wind would exclusively form a strong K$\alpha$ line of He-like Ne and H-like N with P-Cygni profiles in the spectrum taken at $t=t_{1296}$ and eliminate other lines by line blanketing effects. The line blanketing occurs when the Doppler broadened line profile overlaps with that of another line at a higher energy or shorter wavelength, and weakens the latter line and strengthen the former line. A supersonic wind is expected to enhance this effect owing to large Doppler shifts of the line energies. We use a Monte Carlo method to calculate the photon transfer and try to reproduce the spectral features at $t=t_{1296}$.

 In Section \ref{sec:formulation}, we present our simplified stationary wind model including the velocity and density profiles and describe the procedure to determine the mass loss rate from observed quantities. The radiative processes involved in our Monte Carlo method are presented in Section \ref{sec:MCcode}.  In Section \ref{sec:results}, we show the results and compare them with observations. We conclude this paper in Section \ref{sec:conclusions}.

\section{Model of Accelerating Wind}\label{sec:formulation}
To reproduce the spectrum at $t=t_{1296}$, we construct a model of nova wind by solving simplified equations for a spherically symmetric wind above the photosphere. Because of the super-Eddington luminosity of MAXI J0158$-$744, the gas pressure is significantly smaller than the radiation pressure. Therefore the equation of motion can be approximated as
\begin{eqnarray}
v\frac{dv}{dr}=\frac{\kappa L}{4\pi r^{2}c}\left(1-\frac{L_{\rm Edd}}{L}\right)\label{eq:equation-motion01},
\end{eqnarray}
where $v$ is the velocity, $r$ the radial distance from the center, $\kappa$ the opacity, $L$ the luminosity,  $c$ the speed of light. The Eddington luminosity  $L_{\rm Edd}$ of a WD with the mass $M_{\rm WD}$ has been introduced as $L_{\rm Edd}=4\pi cGM_{\rm WD}/\kappa$, where $G$ denotes the gravitational constant. Here we have assumed that the wind becomes stationary, because the time $t_{1296}$ is much longer than the dynamical timescale of this wind, that is $t_{\rm d}\sim R_{1296}/v=1-0.1$ s for the typical wind velocity of novae ($v\sim10^{3}-10^{4}$ km s$^{-1}$; \citealt{2011ApJS..197...31S}). The solution of Equation (\ref{eq:equation-motion01}) is expressed as
\begin{eqnarray}
v=\sqrt{\frac{\kappa L}{2\pi cR_{0}}\left(1-\frac{L_{\rm Edd}}{L}\right)\left(1-\frac{R_{0}}{r}\right)+v_{0}^{2}}\label{eq:velocity01},
\end{eqnarray}
where $R_{0}$ is the photospheric radius, which is equal to $R_{1296}$ at $t=t_{1296}$, and $v_{0}$ is the velocity at $r=R_{0}$. Note that $L_{\rm Edd}/L$ is of the order of 0.01. The velocity at $r\rightarrow\infty$ approaches $v_{\rm max}\sim\sqrt{\kappa L/2\pi cR_{0}}$ if $v_{0}<<v_{\rm max}$. Here we have assumed that the luminosity $L$ is kept to be constant at $L_{1296}$ above the photosphere. The initial velocity $v_{0}$ needs to be significantly smaller than the maximum velocity $v_{\rm max}$ for the line blanketing to work efficiently. In the following, we set $v_{0}=100$ km s$^{-1}$. 

The mass loss rate is determined so that the optical depth of the solution with respect to the planck mean opacity be 2/3. To do this, we need to calculate the ionization states in the wind. The mass loss rate in turn affects the ionization states in the wind through the ionization parameter for a photo-ionized plasma. Therefore we need to iterate this procedure to obtain an appropriate mass loss rate.

\subsection{Ionization States} 
\begin{figure}[htp]
\epsscale{0.6}
\plotone{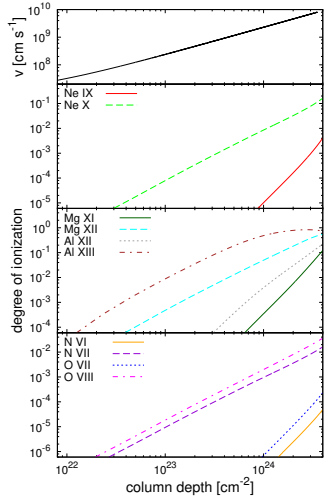}
\caption{Distributions of expansion velocity (top panel) and degree of ionization (i.e., the ratio of ion density in each ionized state to the total density) for each element in the wind. The column depth is measured from the photosphere.}
\label{fig:outvar}
\end{figure}

We use the XSTAR subroutines  \citep{2001ApJS..133..221K} to calculate ionization states in the wind  irradiated by the blackbody radiation with the temperature of $T=T_{1296}$ and the photospheric radius of $R_{0}=R_{1296}$.  The bolometric luminosity becomes $8.0\times10^{39}$ erg s$^{-1}$. Though the actual wind solution has a radial density profile, we treat the wind as a gas sphere with a constant density that reproduces the same column depth for simplicity. This assumption overestimates the  ionization parameters near the photosphere and underestimates them in the outer layer.  In addition, we do not include the adiabatic cooling in the code, which may dominate the cooling process in the supersonic wind. Instead, we assume a gas temperature (33 eV) significantly smaller than the photospheric temperature to use the XSTAR with minimum modifications. Thus we obtain the ionization states of relevant elements as functions of the column depth measured from the photosphere (Figure \ref{fig:outvar}). Assigning the ionization states to positions in the wind with the same column depth, we obtain the ionization states in the wind. We leave a detailed and self-consistent treatment of ionization states in supersonic winds as a future work.

The mass fraction of the heavy elements except for CNO and Ne is assumed to be 0.1 solar value (see Table \ref{tb:massfraction-smc}; the solar abundance is taken from \citet{1989GeCoA..53..197A}) in the wind to mimic the SMC abundance \citep{2008AJ....136.1039C}. The compositions of CNO elements should be modified by the CNO cycle operating beneath the photosphere. We treat the Ne mass fraction as a free parameter to reproduce the observed strong Ne line. We will discuss this issue in Section \ref{sec:results}.
%\item{constant density}
%\item{ion distribution with respect to column depth}
%\item{mass fraction=SMC=0.1 solar}
%\item{oxygen changed to reproduce the observed spectrum}

\begin{table}[h]
\begin{center}
\begin{tabular}{|c|c|}
\hline
&mass fraction\\ \hline
N&$1.09\times10^{-4}$\\ \hline
O&$1.02\times10^{-3}$\\ \hline
Ne&$1.85\times10^{-4}$\\ \hline
Mg&$6.93\times10^{-5}$\\ \hline
Al&$6.20\times10^{-6}$\\ \hline
\end{tabular}
\end{center}
\caption{Reference value for the mass fraction of each element, corresponding to 10\% of the solar abundance, a typical SMC abundance \citep{1989GeCoA..53..197A}.}
\label{tb:massfraction-smc}
\end{table}

\section{Monte Carlo Method}\label{sec:MCcode}
Resonance line scattering of photons in a spherically expanding wind would form P-Cygni profiles on the spectrum and may cause line blanketing effects depending on the velocity distribution in the wind.
We calculate the transfer of photons in the expanding wind by a Monte Carlo simulation.
Here we describe the boundary conditions and the treatment of radiative processes involved in the simulation.

\subsection{Transfer of Photons}
%\subsection{Distribution of Photons}
 The calculation starts with photons isotropically emitted at the photosphere (at which the radius is $R_{1296}$ and the optical depth equals 2/3). Their energy distribution is expressed by the Planck function with the temperature of $T_{0}$ in the rest frame of the wind, i.e.,
\begin{eqnarray}
B_{{\rm ph},\epsilon^\prime}=\frac{2}{h^{3}c^{2}}\frac{{\epsilon^\prime}^{2}}{\exp(\epsilon^\prime/k_{\rm B}T_{0})-1},
\end{eqnarray}
where $\epsilon^\prime$ denotes the photon energy in the rest frame of the wind, $h$ is the Planck constant. We adopted $T_{0}=T_{1296}$ suggested by the MAXI observation. %We specify the direction of the momentum of a photon with the inclination angle $\theta$ and the azimuth angular $\phi$.

%We calculate the thermalization length according to the formula 
%\begin{equation}
%\tau_{*}=\int_r^\infty\sqrt{(\kappa\rho+\sum_{\rm i}n_{\rm i}\sigma_{\rm bf})\sum_{\rm i}n_{\rm i}\sigma_{\rm bf}}dr,
%\end{equation}
 %for each photon, and generate it at the radius where $\tau_{*}$ becomes 2/3.

%\textcolor{red}{ground state. $\rho\sim\dot{M}/4\pi R^{2}v\sim10^{-5}$ g/cc, $n_{\rm e}\sim\rho/m_{\rm H}\sim10^{19}$ /cc, $\sigma n_{\rm e}v\sim10^{10}$ /s.}

%\textcolor{red}{In the case of Ne IX, $A_{21}\sim8\times10^{12}$/s}.

Seed photons with the energy of $\epsilon$ emitted from the photosphere can be absorbed by ions with the velocity $v$, only when the Doppler shifted energy $\epsilon^{\prime}$ of the photons in the rest frame of the ions match precisely the bound-bound transition energy $\epsilon_{\rm lu}$ of the ions, from the lower ($n={\rm l}$) to upper ($n={\rm u}$) state,
\begin{eqnarray}
\epsilon_{\rm lu}=\epsilon\gamma\left[1-(v/c)\cos{\varphi}\right].\label{eq:Doppler}
\end{eqnarray}

Here $\epsilon$ is the photon energy in the fixed frame, $\gamma=1/\sqrt{1-(v/c)^2}$ is the Lorentz factor, $\varphi$  the angle between the momentum of a photon and the radial direction. 
Then the total cross section as a function of the photon energy $\epsilon^\prime$ in the ion rest frame is expressed as the Lorentz profile
\begin{eqnarray}\label{eq:crss}
\sigma(\epsilon^\prime)=\frac{\pi r_{0}c\hbar^{2}A_{\rm ul}}{(\epsilon^\prime-\epsilon_{\rm lu})^{2}+(\hbar A_{\rm ul}/2)^{2}}\label{eq:crds},
\end{eqnarray}
where $\hbar$ denotes the Dirac constant and the Einstein  $A$ coefficient is written by
\begin{eqnarray}
A_{\rm ul}=\frac{1}{g_{\rm u}}\frac{8\pi^{2}r_{0}\epsilon_{\rm lu}^{2}}{ch^{2}}g_{\rm l}f_{\rm lu},
\end{eqnarray}
where $g_{\rm u}$ and $g_{\rm l}$ are the statistical weights of the upper and lower states, respectively, $r_{0}$ is the classical electron radius, and $f_{\rm lu}$ the oscillator strength \citep{1979rpa..book.....R}.
 Since the spontaneous transition from the state $n={\rm u}$ to the ground state is much faster than the reverse transition, we assume all ions are in the ground state in the transfer calculations. The fast spontaneous transition also defeats collisional de-excitations. Thus the whole process of  absorption by exciting an ion followed by emission due to the reverse transition is recognized as scattering.

Using the cross section in Equation (\ref{eq:crds}) we calculate the probability $P$ of scattering off an ion i expressed as $1-\exp\left[-\int^{l_{\rm ed}}_{l_{\rm st}}\sigma(\epsilon^{\prime})n_{\rm i}dl\right]$, where $n_{\rm i}$ is the number density of the ions i and the integration is performed through the path of a photon from $l_{\rm st}$ to $l_{\rm ed}$. Since $dl$ is converted to the energy shift in the rest frame of the ion by $dr/\cos\varphi=(dr/dv)(dv/d\epsilon^{\prime})(d\epsilon^{\prime}/\cos\varphi)$, where $dr/dv$ and $dv/d\epsilon^{\prime}$is derived from Equation (\ref{eq:equation-motion01}) and the Doppler effect equation (\ref{eq:Doppler}), $P$ is replaced by
\begin{eqnarray}
P=1-\exp\left[-\int^{\epsilon^{\prime}_{\rm ed}}_{\epsilon^{\prime}_{\rm st}}\sigma(\epsilon^{\prime})n_{\rm i}\frac{dr}{dv}\frac{dv}{d\epsilon^{\prime}}\frac{d\epsilon^{\prime}}{\cos\varphi}\right].
\end{eqnarray}
Note that $\epsilon^{\prime}_{\rm st}$ is larger than $\epsilon^{\prime}_{\rm ed}$, while $l_{\rm st}$ is smaller than $l_{\rm ed}$. Here we approximate the integration of $\sigma(\epsilon^{\prime})d\epsilon^{\prime}$ by $\sigma(\epsilon_{\rm lu})\Delta\epsilon^{\prime}$, where $|\Delta\epsilon^{\prime}|=(v_{\rm t}/c)\epsilon_{\rm lu}$ denotes the line width due to thermal Doppler broadening (\citealt{sobolev1960moving}). Therefore $P$ can be written by
\begin{eqnarray}
P=1-\exp\left[-\sigma(\epsilon_{\rm lu})n_{\rm i}\frac{dr}{dv}\frac{1-(v/c)\cos\varphi}{\cos^{2}\varphi}v_{\rm t}\right],
\end{eqnarray}
in the non-relativistic case.

Each ion follows the Maxwell-Boltzmann distribution in the rest frame of the wind, 
\begin{eqnarray}
F_{\rm i}\propto p_{\rm i}^{2}\exp\left[-\frac{p_{\rm i}^{2}}{2m_{\rm i}k_{\rm B}T_{\rm i}}\right],
\end{eqnarray}
where $p_{\rm i}$ is the ion momentum, $m_{\rm i}$ is the ion mass, and $T_{\rm i}$ is the ion temperature. 

The energy of the scattered photon  is determined so that the energy distribution follows the Lorentz profile of Equation (\ref{eq:crss}). Each photon repeats the scattering process described in this section until $t=t_{1296}$, unless it turns back into the photosphere.

In this study,  we take into account 9 transitions from the ground states  for He-like and H-like ions of N, O, Ne, Mg, and Al, i.e., the upper states are 1s2p$^{1}$P$_{0}$ to 1s10p. Table \ref{tb:transition01} lists the first two transitions of each ion included in our calculations.

\begin{table}[htb]
\begin{center}
\begin{tabular}{|c|c|c|c|c|}
\hline
&\multirow{2}{*}{\shortstack{ionization\\state}}&\multirow{2}{*}{\shortstack{upper\\state}}&\multirow{2}{*}{\shortstack{energy\\ {[keV]}}}&\multirow{2}{*}{\shortstack{oscillator\\strength}}\\[4mm] \hline
\multirow{4}{*}{O}&\multirow{2}{*}{He-like}&1s2p&0.57&0.69\\ \cline{3-5}
&&1s3p&0.67&0.15\\ \cline{2-5}
&\multirow{2}{*}{H-like}&1s2p&0.65&0.42\\ \cline{3-5}
&&1s3p&0.77&0.08\\ \hline
\multirow{4}{*}{Ne}&\multirow{2}{*}{He-like}&1s2p&0.92&0.72\\ \cline{3-5}
&&1s3p&1.07&0.15\\ \cline{2-5}
&\multirow{2}{*}{H-like}&1s2p&1.02&0.42\\ \cline{3-5}
&&1s3p&1.21&0.08\\ \hline
\multirow{4}{*}{Mg}&\multirow{2}{*}{He-like}&1s2p&1.35&0.74\\ \cline{3-5}
&&1s3p&1.58&0.15\\ \cline{2-5}
&\multirow{2}{*}{H-like}&1s2p&1.47&0.42\\ \cline{3-5}
&&1s3p&1.74&0.08\\ \hline
\multirow{4}{*}{Al}&\multirow{2}{*}{He-like}&1s2p&1.60&0.75\\ \cline{3-5}
&&1s3p&1.87&0.15\\ \cline{2-5}
&\multirow{2}{*}{H-like}&1s2p&1.72&0.42\\ \cline{3-5}
&&1s3p&2.04&0.08\\ \hline
\end{tabular}
\end{center}
\caption{A list of representative transitions adopted in our calculations. The values are fetched from the atomic line list version 2.04 of Peter van Hoof (see http://www.pa.uky.edu/$\sim$peter/atomic/).}
\label{tb:transition01}
\end{table}

\section{Results}\label{sec:results}
\begin{figure}[htp]
\epsscale{0.8}
\plotone{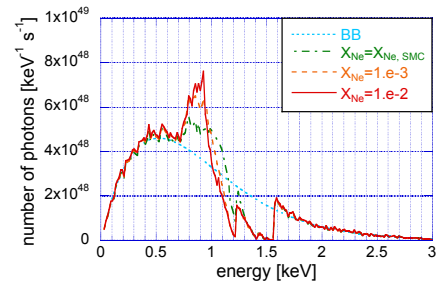}
\caption{Spectra as results of Monte Carlo simulations taking into account only scattering by Ne with different mass fractions. The cyan dotted line is the incident irradiating emission represented by a blackbody spectrum at temperature $T=T_{1296}$. The other three lines represent the spectra with $X_{\rm Ne}=1.85\times10^{-4}$ (the same as the SMC abundance; green dash-dotted line),   $X_{\rm Ne}=1\times10^{-3}$ (the orange dashed line), and  $X_{\rm Ne}=10^{-2}$ (the red solid line).}
\label{fig:neix-line}
\end{figure}

First, we present results of our calculation taking only Ne lines into account and derive the required amount of Ne to produce a distinctive peak due to the  K$\alpha$ line of Ne IX. Figure \ref{fig:neix-line} displays the dependence of the shape of the Ne line on the mass fraction, where $X_{\rm Ne}=1.85\times10^{-4}$ (green dash-dotted line), $1.0\times10^{-3}$ (orange dashed line), and $10^{-2}$ (red solid line). The cyan dotted line represents the blackbody continuum with the temperature of $T_{1296}$ emitted from the photosphere at the radius of $R_{1296}$. When the Ne mass fraction is less than $10^{-3}$, a broad Ne line is formed and the peak appears at an energy higher than 0.92 keV (the transition energy of Ne IX K$\alpha$ in the rest frame), which is incompatible with
the MAXI observation.  This is because Ne IX K$\alpha$ line becomes optically thin with this small Ne mass fraction while Ne X K$\alpha$ line is still optically thick. Instead, when Ne IX K$\alpha$ line is optically thick due to the large Ne mass fraction, photons emitted by Ne X K$\alpha$ line at the energy of 1.02 keV in the rest frame of the ions are scattered by K$\alpha$ transitions of Ne IX ions because the energy of the photons is reduced by as much as $\sim17$ \% and can become exactly 0.92 keV in the rest frame of some Ne IX ions in the surrounding matter receding  at velocities of the order of $v_{\rm max}\sim50,000$ km s$^{-1}$. This scattering off a Ne IX ion degrades the photon energies lower than 0.92 keV in the rest frames of the other ions. As a consequence, the existence of Ne X ions does not form an emission line at 1.02 keV but enhances the K$\alpha$ line of  Ne IX ions. This is the so-called line blanketing effect. Therefore we assume $X_{\rm Ne}=1\times10^{-2}$ in the following calculations so that a distinctive Ne peak surely appears at 0.92 keV. The mass fraction of  Ne ions in the lower ionization stages is too low to affect the spectrum.

\begin{figure}[htp]
\epsscale{0.6}
\plotone{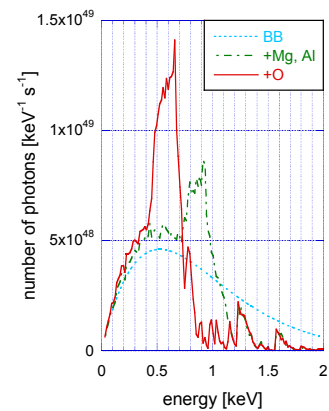}
%\plotone{bfvar-2x1-5kL64e38Ml12e19ox5e-9plm-7-j.eps}
\caption{Comparison of spectra taking account of scattering off other ions as well as Ne. The cyan dotted line is the incident irradiating emission represented by a blackbody spectrum at temperature $T=T_{1296}$. The green dash-dotted line includes  Mg XI, XII \& Al XII, XIII and the red solid line represents the spectrum including O VII and VIII in addition to these ions.}
\label{fig:spectra-var}
\end{figure}

\begin{figure}[htp]
\epsscale{0.6}
\plotone{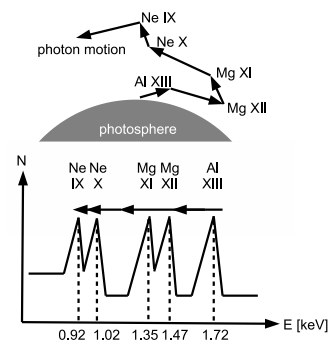}
\caption{A schematic image of a representative photon track, which is scattered several times by different species of ions (upper illustration), and the spectral deformation (line blanketing effects; lower illustration).}
\label{fig:schematic-lineblanketing}
\end{figure}

\begin{figure}[htp]
\epsscale{0.6}
\plotone{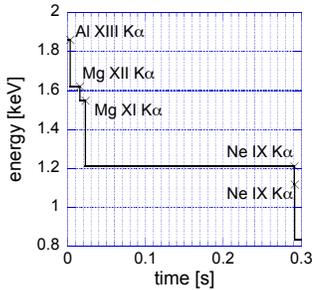}
\caption{An example of photon energy shifts. The initial photon energy of 1.86 keV decreases down to 0.83 keV after scattering sequentially by several ions with transition energies in decreasing order, i.e.,  Al XIII K$\alpha$, Mg XII K$\alpha$, Mg XI K$\alpha$, and Ne IX K$\alpha$.}
\label{fig:ex-Et}
\end{figure}

Next, we investigate effects of other abundant ions, i.e.,  Mg XI, XII, and Al XII, XIII. These ions have transitions with energies higher than the energy of the Ne X K$\alpha$ line 1.02 keV. The green dash-dotted line in Figure \ref{fig:spectra-var} designates the result. The emission lines associated with transitions of Mg and Al ions look rather weak, and it is remarkable that the existence of these elements enhances the Ne line due to line blanketing (This enhancement is verified by comparison with the red solid line in Figure \ref{fig:neix-line}).
Figure \ref{fig:schematic-lineblanketing} illustrates a schematic view of the typical motion of scattered photons. The upper illustration is a schematic image of photons traveling in the wind and scattered by different species of ions in  order of decreasing transition energy such as Al XIII$\to$Mg XII$\to$Mg XI$\to$Ne X$\to$Ne IX. The lower illustration shows how this series of scattering changes the spectrum. Because the photon reduces its energy, the emission lines at energies higher than 0.92 keV become weak. Figure \ref{fig:ex-Et} displays an example of decreasing energy of a photon with time, extracted from our calculation. The energy of a sample photon gradually changes from 1.86 keV
to 1.62, 1.55, 1.21, 1.12 and 0.83 keV due to the scattering off 
Al XIII, Mg XII, Mg XI, Ne IX and Ne IX ions, respectively.

\subsection{Influence of Oxygen Mass Fraction on Ne line}\label{sec:spe-ox-abund}
In contrast to the enhancement of the Ne line by the existence of Mg and Al ions, the inclusion of O VII and VIII ions with the mass fraction of 0.1 solar, which have transitions with the energies lower than 0.92 keV, significantly weakens the K$\alpha$ line of Ne IX and produces a very strong emission line at $\sim0.7$ keV as shown by the red solid line in Figure \ref{fig:spectra-var}.

\begin{figure}[htp]
\epsscale{0.6}
\plotone{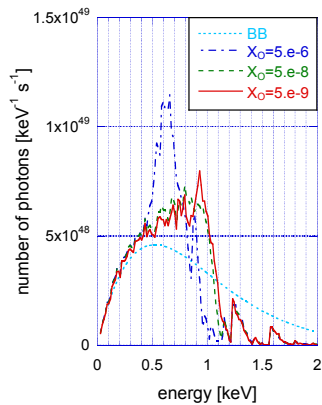}
\caption{Influence of oxygen mass fraction on spectra, i.e., $X_{\rm O}=5\times10^{-6}$ (blue dash-dotted line), $5\times10^{-8}$ (green dashed line), and $5\times10^{-9}$ (red solid line). The cyan dotted line is the incident irradiating emission represented by a blackbody spectrum at temperature $T=T_{1296}$.}
\label{fig:oxygen-influence}
\end{figure}

To reproduce the observed spectrum, the reduction of the oxygen mass fraction is necessary. As a next step, we calculate spectra with reduced oxygen mass fractions and the results are shown in Figure \ref{fig:oxygen-influence}. Here, the influence on the Ne line profile becomes small when the O mass fraction is between $X_{\rm O}=5\times10^{-8}$ (green dotted line) and $5\times10^{-9}$ (red solid line). 
%The equivalent widths of the line is estimated to be  0.25 keV ($X_{\rm O}$=$5\times10^{-8}$) and 0.27 keV ($X_{\rm O}$=$5\times10^{-9}$).
If $X_{\rm O}=5\times10^{-8}$ then the line at 0.92 keV disappears and the K$\alpha$ line of O VIII alternatively becomes prominent. So, we refer to the model with $X_{\rm O}$=$5\times10^{-9}$ as the fiducial model. The fiducial model has the mass loss rate of $\approx9.1\times10^{19}$ g s$^{-1}$.

\subsection{Oxygen Depletion in the Wind}
\begin{figure}[htp]
\epsscale{0.9}
\plotone{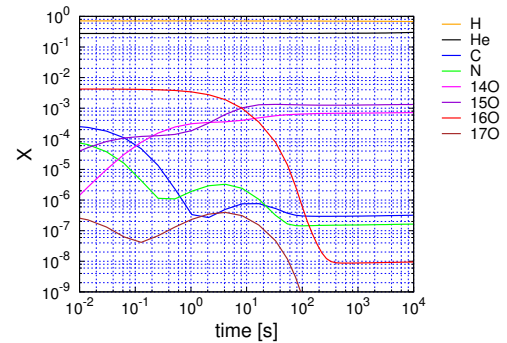}
\caption{The evolution of the mass fraction of CNO elements. The CNO cycle reaction starts at $t=0$. Most of $^{16}$O disappears in $\sim600$ s (from $X_{\rm O}=1\times10^{-3}$ to $4\times10^{-9}$), and replaced by $^{15}$O and $^{14}$O. $^{15}$O ($^{14}$O) would decay into $^{15}$N ($^{14}$N) with the half-life time of 122 s (71 s) after the end of the CNO cycle.}
\label{fig:abhist-j}
\end{figure}

The reduction of the O mass fraction required to reproduce the observed spectrum can be realized by the CNO cycle. To demonstrate this, we perform a nuclear reaction network calculation for a gas with a fixed temperature ($3\times10^8$ K) and density (100 g cm$^{-3}$) using the code developed by one of the authors \citep{2010AIPC.1279..415S}. The initial composition of heavy elements is 0.1 solar with the enhanced O and Ne mass fractions. The mass fraction of Ne is $X_{\rm Ne}=0.01$. The corresponding mass fraction of O becomes 0.4 solar ($X_{\rm O}=4.1\times 10^{-3}$) if we estimate from the surface composition of an ONe WD model with the mass of 1.37 $M_\odot$ \citep{1994ApJ...434..306G}. According to this modeling, the mass fraction of O on the surface of the core decreases with increasing stellar mass. Thus a WD with the mass closer to the Chandrasekhar limit leads to a smaller $X_{\rm O}$. Figure \ref{fig:abhist-j} shows the resultant temporal evolution of the composition of elements involved in the CNO cycle.   Most of the initial CNO elements are converted to unstable isotopes $^{15}$O and $^{14}$O in $\sim600$ s. Since the half life of  $^{15}$O ($^{14}$O) is 122 s (71 s),  the required reduction of the O mass fraction suggests that the CNO cycle continued only for $\sim10^3$ s and then the  synthesized $^{15}$O ($^{14}$O) decayed to $^{15}$N ($^{14}$N) for the subsequent $\sim$2400 s or more. 

In other words, we need an interval of more than $\sim$2400 s after the termination of the CNO cycle to eliminate O  in the wind and to form a strong Ne line in the spectrum at $t=t_{1296}$. Then the CNO cycle must have stopped before $t\sim t_{1296}-2400\;{\rm s}=-1100$ s. No detection of the event by MAXI at $t=-5530$ s ($=t_{-5530}$) further constrains the duration of the CNO cycle  ($\Delta t_{\rm CNO}$), that is $\Delta t_{\rm CNO}<-1100\;{\rm s}-t_{-5530}=4430$ s. 

The resultant O mass fraction is two times higher than the required value derived in the previous section and the spectrum with this higher O mass fraction has the peak of the Ne IX line at a slightly higher energy (0.95 keV) and marginally consistent with the observation. This might indicate that the WD is more massive than 1.37 $M_\odot$ and has a mass closer to the Chandrasekhar limit.
%This timescale becomes shorter by $\sim$15\% if the mass fraction of $^{15}$O in the wind is enhanced by a factor similar to Ne.

\subsection{Nitrogen}\label{sec:nitrogen}
Before finally coming to the conclusions, it is indispensable to examine whether N ions increased through the O decay argued in the former section is compatible with the observed Ne line or not. We calculate a spectrum including N VI and VII, assuming the same mass fraction of N as the initial $X_{\rm O}$ ($=X_{\rm O,\,SMC}=1\times10^{-3}$). A list of mass fractions of elements we adopt in this calculation is shown in Table \ref{tb:massfraction-model}.

\begin{figure}[htp]
\epsscale{0.6}
\plotone{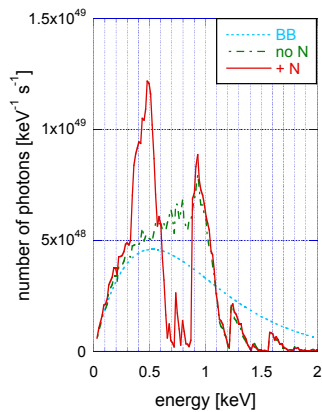}
\caption{Comparison of spectra with  (red solid line) and without (green dash-dotted line) N VI and VII transitions (The cyan dotted line is the incident irradiating emission represented by a blackbody spectrum at temperature $T=T_{1296}$.). These transitions scatter photons with the energy less than 0.87 keV, which should be recognized as a part of a broaden Ne line in the latter spectrum. The half width and the equivalent width of the Ne line (red solid line) are 0.080 keV and 0.18 keV, respectively.}
\label{fig:nitrogen-influence}
\end{figure}

The result is shown by the red solid line in Figure \ref{fig:nitrogen-influence}. In comparison with the calculation of "no N" (green dash-dotted line), the existence of N sharpens the shape of the Ne line. It is again caused by line blanketing due to N VI and VII ions. In this calculation, most photons forming the lower energy part of the Ne line ($\lesssim0.87$ keV) are scattered by N ions and converted to N line at lower energies. Here, the half width at 1/e times the Ne peak above the blackbody heights is 0.080 keV, and the equivalent width is 0.18 keV, which  are compatible with the observations \citep{2013ApJ...779..118M}. 

A strong N line appears at $\sim0.5$ keV close to the lower limit of the energy range covered by the MAXI/SSC. This must be compatible with the observation because the sensitivity at 0.5 keV is a factor of about 3 weaker than that at 1 keV.

%\begin{figure}[htp]
%\epsscale{0.4}
%\plotone{bfx22-7k-L1-2e39ne1e-2.eps}
%\caption{Spectral deformation associated with the luminosity. When the luminosity is below $1\times10^{39}$ erg s$^{-1}$, the effect of line blanketing could not enhance the Ne line.}
%\label{fig:L-lineblanketing}
%\end{figure}

\begin{table}[htp]
\begin{center}
\begin{tabular}{|c|c|}
\hline
&mass fraction\\ \hline
N&$1.02\times10^{-3}$\\ \hline
O&$5.0\times10^{-9}$\\ \hline
Ne&$1.0\times10^{-2}$\\ \hline
Mg&$6.93\times10^{-5}$\\ \hline
Al&$6.20\times10^{-6}$\\ \hline
\end{tabular}
\end{center}
\caption{Mass fraction of each element assumed in the fiducial model.  See text for details.}
\label{tb:massfraction-model}
\end{table}

\subsection{Influence of the Luminosity on Spectrum}
\begin{figure}[htp]
\epsscale{0.6}
\plotone{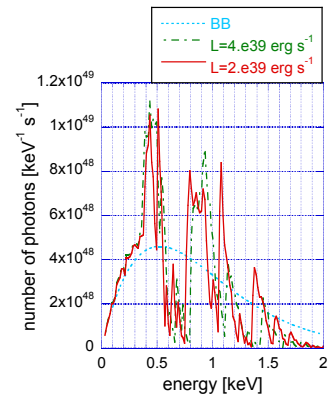}
\caption{Spectra calculated for two values of the luminosity in Equation (\ref{eq:velocity01}) ($4\times10^{39}$ erg s$^{-1}$ for the green dash-dotted line and $2\times10^{39}$ erg s$^{-1}$ for the red solid line. The cyan dotted line is the blackbody continuum for $T=T_{1296}$.).}
\label{fig:luminosity-influence}
\end{figure}

Because a luminosity derived from a blackbody fit to an observed spectrum has a large uncertainty in general \citep{1996ApJ...456..788K}, we investigate how the luminosity affects the line blanketing. To do this, we decrease the value of luminosity in Equation (\ref{eq:velocity01}) from that used in the former section to calculate ion scattering in the wind having lower velocities, while keeping the other parameter values such as the photospheric radius and the temperature unchanged as in Section \ref{sec:nitrogen}. Figure \ref{fig:luminosity-influence} shows the results. For $L=4\times10^{39}$ erg s$^{-1}$, a prominent peak appears at Ne X K$\alpha$ energy (the green dash-dotted line). For $L=2\times10^{39}$ erg s$^{-1}$, the peak shifts to a lower energy. Therefore a luminosity brighter than  $4\times10^{39}$ erg s$^{-1}$ is required to enhance Ne IX K$\alpha$ line by line blanketing.
%Because line blanketing is caused by the Doppler effect, super-Eddington luminosity ($\gtrsim1\times10^{39}$ erg s$^{-1}$) is required to efficiently enhance the Ne line.

\section{Conclusions and Discussions}\label{sec:conclusions}
We have examined whether resonance scattering off N,  O, Ne, Mg, and Al ions in an accelerating wind reproduces a strong  Ne IX K$\alpha$  emission line and lack of the Ne X K$\alpha$ line  in  the observed X-ray spectrum of MAXI J0158$-$744, by performing Monte Carlo calculations. We have constructed a simplified stationary model for a given luminosity with the ion distribution calculated using the XSTAR subroutines.  The mass loss rate is determined to reproduce the observed photospheric radius.

As a result of a series of the calculations presented here,  the observed spectral features can be reproduced by emission from the expanding wind irradiated by a super-Eddington soft X-ray source. It is found that the Ne mass fraction needs to be enhanced by a factor of several tens of times compared with the SMC abundance ($X_{\rm Ne}\sim1\times10^{-2}$) to reproduce the observed Ne line at 0.92 keV.  This amount of Ne could be supplied by  matter dredged up from the surface of a WD composed of  ONe.  It should be noted that the ionization states estimated by ignoring the advection in the wind might overestimates the ionization degree and thus the necessary mass fraction of Ne. %We also find that the inclusion of Ne X ions at energies higher than 0.92 keV supplies more photons in the Ne IX K$\alpha$ line due to line blanketing rather than forms the Ne X K$\alpha$ line.
We also find that the inclusion of Ne X ions at energies higher than 0.92 keV does not induce the Ne X K$\alpha$ line due to the line blanketing. In the same way, inclusion of heavier elements like Mg and Al supply more photons to the Ne IX K$\alpha$ line and strengthen it. Meanwhile, even a small amount of O weakens the Ne emission line for the same reason. It is necessary to decrease the mass fraction of O under $5\times10^{-9}$ for removing the influence on the Ne line completely. We have shown that the CNO cycle beneath the photosphere can reduce the amount  of O. To reduce its mass fraction under $5\times10^{-9}$,  the duration of the CNO cycle is constrained to be shorter than 4430 s.

The inferred hydrogen mass burned out by the CNO cycle is $\approx10^{-8}M_{\odot}$. This is far smaller than the minimum accretion mass that enables ignition in hydrostatic equilibrium, $5\times10^{-7}M_{\odot}$  \citep{1982ApJ...257..767F}. Taking this into account, hydrogen should be accreted dynamically onto the surface of the WD and the subsequently generated shock ignites hydrogen leading to TNR. Such a dynamical behavior may be responsible for the dredge up of Ne ions, which is needed to explain the observed Ne line. The mass of such a WD must be very close to the Chandrasekhar limit, which was also indicated from the small  photospheric radii deduced from observations \citep{2013ApJ...779..118M}.

\acknowledgments
TS is grateful to Tatehiro Mihara for giving us an opportunity to think about modeling this very intriguing source and thanks Izumi Hachisu for useful comments.

\bibliographystyle{apj}
\bibliography{hogehoge}

%\begin{thebibliography}{}
%\bibitem[Arnett(1988)]{1988ApJ...331..377A} Arnett, W.~D.\ 1988, \apj, 331, 377
%\end{thebibliography}

\end{document}